\documentclass[useAMS,usenatbib]{mn2e}

\usepackage{amssymb}
\usepackage{amsfonts}
\usepackage{mncite}
\usepackage{epsfig}
\usepackage{psfig}

\defcitealias{LR04}{Paper I}

\begin{document}

\title[Planetesimal formation in self-gravitating discs]{Planetesimal
formation via fragmentation in self-gravitating protoplanetary discs}

\author[W.K.M. Rice, G. Lodato, J.E. Pringle, P.J. Armitage, \& I.A. Bonnell]{W.K.M. Rice$^1$\thanks{E-mail: wkmr@roe.ac.uk},
G. Lodato$^2$, J.E. Pringle$^2$, P.J. Armitage$^{3,4}$ and I.A. Bonnell$^5$\\ 
$^1$ SUPA\thanks{Scottish Universities Physics Alliance}, Institute for Astronomy, University of Edinburgh,
Blackford Hill, Edinburgh EH9 3HJ \\
$^2$ Institute of Astronomy, Madingley Road, Cambridge, CB3 0HA  \\ 
$^3$ JILA, Campus Box 440, University of Colorado, Boulder, CO
80309-0440, USA \\ 
$^4$ Department of Astrophysical and Planetary Sciences, University
of Colorado, Boulder, CO 80309-0391, USA \\
$^5$ SUPA$\dagger$, School of Physics and Astronomy, University of St Andrews, North Haugh,
St Andrews, KY16 9SS}
 
\maketitle

\begin{abstract}
An unsolved issue in the standard core accretion model for gaseous
planet formation is how kilometre-sized planetesimals form from,
initially, micron-sized dust grains. Solid growth beyond metre sizes
can be difficult both because the sticking efficiency becomes very
small, and because these particles should rapidly migrate into the
central star.  We consider here how metre sized particles evolve in
self-gravitating accretion discs using simulations in which the
gravitational influence of the solid particles is also included.
Metre-sized particles become strongly concentrated in the spiral
structures present in the disc and, if the solid to gas density ratio
is sufficiently high, can fragment due to their own self-gravity to
form planetesimals directly.  This result suggests that planetesimal
formation may occur very early in the star formation process while
discs are still massive enough to be self-gravitating. The dependence
of this process on the surface density of the solids is also consistent
with the observation that extrasolar planets are preferentially found
around high metallicity stars.
\end{abstract}


\section{Introduction}
There are currently two models for the formation of gaseous,
Jupiter-like planets.  The most widely accepted is the core accretion
model \citep{pollack96} in which, initially, a core of rock/ice grows
via the collisional accumulation of planetesimals \citep{safronov} and
then, once sufficiently massive, it accretes a gaseous envelope
\citep{lissauer93, pollack96}. The alternative scenario does not
require the initial growth of a core and assumes that protoplanetary
discs may become gravitationally unstable and that gas giant planets
may subsequently form via direct gravitational collapse
\citep{boss98,boss00,mayer04}.

One aspect of the core accretion model that has yet to be
satisfactorily resolved is how the kilometre-sized planetesimals, that
ultimately coagulate to form the planetary core, grow from, initially,
micron-sized dust grains (e.g. \citealt{safronov,weidencuzzi93}). Core
accretion models generally start with the assumption that these
planetesimal have already formed. However, solid growth beyond metre
sizes can be very difficult, due to two effects. On the one hand the
sticking efficiency of solids becomes relatively small in this size
range (e.g., \citealt{supulver97}). On the other hand, it is well known
that solid particles are influenced by gas drag and, for standard disc
geometries, will generally lose angular momentum and migrate in towards
the central star at a rate that depends on the particle size
\citep{weiden77}.  For a circumstellar disc with properties appropriate
for planet formation, the maximum inward radial velocity will generally
occur for particles with sizes between $1$ cm and $1$ m and may easily
exceed $10^3$ cm s$^{-1}$ \citep{weiden77}. With such large inward
radial velocities, it is possible that these objects may drift into the
central star before becoming large enough to decouple from the disc
gas, preventing the growth of the planetesimals required for the
formation of the planetary cores.

The instability model, since it does not require the formation of a
core, is unaffected by the above process.  This scenario does, however,
require that the disc be relatively massive, and that the disc be able
to cool extremely efficiently (e.g. \citealt{gammie01,
rice03a}). Although very few Class II protostars appear to have discs
with the required mass \citep{becksarg91}, recent observations suggest
that massive discs may be present during the Class 0
\citep{rodriguez05} and Class I \citep{eisner05} phases.  Although this
suggests that massive discs may be present at some stage during the
star formation process, it is still not clear that such discs can cool
sufficiently rapidly for the direct formation of gas giant planets.

Even if gas giant planets cannot form via direct gravitational collapse
in discs around Class 0 and Class I protostars, it is still likely that
these discs will experience self-gravitating phases in which spiral
structures will develop \citep{linpringle87}.  The gas pressure on the inner edges of these
spiral waves increases with radius, leading to super-Keplerian gas
velocity. The drag force between the solid particles and the gas
then results in the dust grains and small particles drifting towards 
the peaks of the spiral structures \citep{haghighipour03a}, 
rather than simply drifting
towards the central star.  \citet{rice04} have already shown, using
numerical simulations, that for certain particles sizes, the local
density could be enhanced by a factor of 10 or more.  A similar effect
would occur in the presence of rings \citep{durisen05}, vortices
\citep{godon00,klahr03} and it has also been shown that density
variations in the disc gas resulting from magneto-rotational turbulence
can also produce significant enhancements in particle densities
\citep{johansen06}.

Particle density enhancements may aid planetesimal formation in two
ways.  Firstly, the enhanced collision rate \citep{rice04} could aid
collisional grain growth, and secondly, the densities achieved may be
sufficient for planetesimal formation through direct gravitational
collapse of the solid component of the disc
\citep{goldreichward,youdinshu02}. It is this second possibility that
we investigate in this paper.  We extend the work of \citet{rice04} to
include the gravitational influence of the solid particles. We consider
three-dimensional, global simulations of self-gravitating accretion
discs in which the gas is maintained in a state of marginal
gravitational instability and is coupled to the solid particles via a
drag force.  We include the gravitational influence of both the gas and
the solid particles self-consistently and we consider different
particle sizes and different initial gas to solid particle surface
density ratios.  We find that if the initial particle surface density
is sufficiently high, a significant fraction of the solid component of
the disc, which is strongly concentrated in the self-gravitating spiral
waves, can become unstable and subsequently fragment into bound
objects.  We therefore conclude that what used to be considered the
most difficult step in planetesimal formation (i.e. the growth beyond
metre sizes) could actually be very rapid, thus leading directly to
kilometre sized objects.
\section{Numerical simulations}
\label{numsims}
The simulations performed here are very similar to those of
\citet{rice04}. We consider a system comprising a point mass,
representing the central star, surrounded by two interpenetrating
discs, a gas disc, and a `planetesimal' disc.
\subsection{The gas disc}
The three-dimensional gaseous disc is modelled using smoothed particle
hydrodynamics (SPH), a Lagrangian hydrodynamics code \citep{benz90,
monaghan92}, and is represented by 250000 pseudo-particles.  Each
particle has a mass, an internal energy and a smoothing length that
varies with time to ensure that the number of neighbours (SPH particles
within two smoothing lenghts) remains $\sim 50$. These neighbouring
particles are used to determine the gas density which, together with
the internal energy, is used to determine the gas pressure. A tree is
used to determine gravitational forces, and to determine the gas
particle neighbours.

The simulation starts by considering only the gaseous disc which has a
mass $M_{\rm disc} = 0.25$ M$_\odot$ and extends from $0.25$ to $25$ au
around a $1$ M$_\odot$ central star. The initial surface density
profile is $\Sigma_{\rm gas} \propto R^{-1}$, the initial temperature
profile is $T \propto R^{-0.5}$, and the temperature is normalised such
that the disc is initially gravitationally stable at all radii
\citep{toomre64}. The disc gas is allowed to heat up through both $P$dV
work and viscous dissipation with the viscosity given by the standard
SPH viscosity \citep{monaghan92}. In the absence of cooling the gas has
an adiabatic equation of state with $\gamma = 5/3$.

The disc gas is cooled with a radially dependent cooling time of
$t_{\rm cool} = \beta \Omega^{-1}$ where $\Omega$ is the orbital
angular frequency, and $\beta$ is a constant that determines the
cooling rate relative to the angular frequency.  In thermal
equilibrium, the cooling time and the viscous $\alpha$ parameter
\citep{shakura73}, that characterises angular momentum transport, are
related through
\begin{equation}
\alpha = \frac{4}{9 \gamma (\gamma - 1)} \frac{1}{t_{\rm cool} \Omega}.
\label{alpha}
\end{equation}
where $\gamma$ is the ratio of the specific heats.  It has been shown
\citep{rice05} that if the value of $\alpha$ required for thermal
equilibrium is too large ($\alpha > \sim 0.06$) the disc will fragment
rather than settle into a quasi-steady, long-lived state.  We therefore
use $\beta = 7.5$ which ensures that the related $\alpha$ value is
small enough for the disc, in the simulations presented here, to settle
into a long-lived, quasi-steady, self-gravitating state
\citep{gammie01, rice03a, LR04, LR05,rice05}. The `planetesimal' disc
is added to the simulation once this long-lived state has been
achieved.
\subsection{The `planetesimal' disc}
The `planetesimal' disc is also modelled using SPH and is represented
by 125000 pseudo-particles.  It is, however, assumed to be pressureless
and hence the solid particles have no internal energy and experience
only gravitational forces, and a drag force due to the difference
between their velocity and the velocity of the disc gas
\citep{weiden77}.  As discussed in detail in \citet{rice04}, the drag
force depends on the particle size, the local gas density, the local
gas velocity, and the local gas sound speed.  To determine the drag
force each `planetesimal' particle has a `smoothing length' that is
varied to ensure that the number of gas particle neighbours remains
$\sim 50$. These neighbouring particles are then used to calculate the
gas density, velocity, and internal energy at the location of every
`planetesimal' particle using the standard SPH formalism
\citep{monaghan92}. The exact value of the drag force is then
determined by specifying the size of the solid particle. In each
simulation, the `planetesimal' disc is assumed to contain particles of
a single size.

The primary difference between the work presented here and that
presented by \citet{rice04} is that we include here the gravitational
influence of the solid particles, rather than treating them as test
particles.  As with the gas particles, the gravitational force due to
the `planetesimal' particles is determined by including them in the
tree.  The details of the tree method can be found in \citet{benz90},
but essentially distant particles are grouped into nodes and their
gravitational force is computed from the position of the node and its
quadrupole correction. Very nearby particles are included in the
neighbour list of the particle being considered, and the force due to
these neighbouring particles is calculated directly.  A direct
gravitational force calculation that includes a `planetesimal' particle
is also softened.  This is needed because `planetesimal' particles do
not feel any pressure forces and hence may undergo extremely close
encounters with other particles. In most of the simulations we used a
softening of $10^{-2}$ au, but repeated some simulations using
$10^{-1}$ au and $10^{-3}$ au.

The `planetesimal' disc is only introduced once the gas disc has
settled into a long-lived, quasi-steady, self-gravitating state.  In this
self-regulated state, the gravitational stability parameter, $Q \sim 1$,
and the disc temperature increases with radius from $\sim 30$ K in the 
inner disc to $\sim 150$ K in the outer disc. The disc temperature is therefore
well below the dust sublimation temperature of $\sim 1600$ K and we would not expect
grain evaporation to be an important process in the self-gravitating regions of the disc.
It should, however, be noted that the magnitude of the temperature and the
temperature profile are largely determined by our choice of disc mass and
surface density profile. However, it would require a disc almost two orders of
magnitude more massive than the one in these simulations for grain evaporation 
to be an important effect.  A slightly more massive disc could, however, have a
temperature that may be high enough to influence the condensation of ices.

The `planetesimal' disc is initially located only in the midplane ($z = 0$)  and extends from $2$
to $20$ au.  The particles are distributed randomly in such a way as to
give the initial surface density profile of $\Sigma_{\rm pl} \propto
R^{-1}$, the same as that of the initial gas disc. An initial spiral
structure is not introduced into the `planetesimal' disc. Although the
disc is initially located only in the midplane, it is given an
isotropic velocity dispersion with a magnitude of $0.1 c_s$ where $c_s$
is the local gas sound speed.  This ensures that the planetesimal disc
is, at the start of the simulation, gravitationally stable.  The
`planetesimal' particles are free to move vertically allowing the disc
to achieve a finite thickness, which turns out to be comparable to the
gas disc thickness. 

Once the `planetesimals' have been added, the simulation is evolved
for about an additional outer rotation period ($\sim 125$ yrs).  We
consider particle sizes of $150$ cm and $1500$ cm, and gas to dust
surface density ratios of $100$ and $1000$.
\section{Simulation results}
\label{simres}
\begin{figure}
\centerline{\psfig{figure=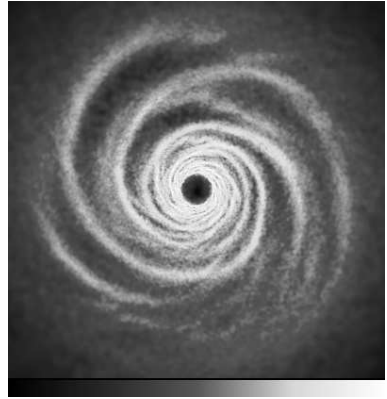, width=50mm}}
\caption{Surface density structure of a $0.25$ M$_\odot$ disc around a
$1$ M$_\odot$ star that has reached a long-lived, self-gravitating
state. The scale shows the logarithm of the surface density,
$\Sigma_{\rm gas}$, with the scale covering $2.4 < {\rm
log}(\Sigma_{\rm gas} / {\rm g cm}^{-2}) < 4.4$.}
\label{gasdisc}
\end{figure}

As discussed above, the simulation initially considers only the gaseous
disc component.  This is evolved until it settles into a marginally
stable, self-gravitating state \citep{LR04}. In this quasi-steady
state, the instability produces spiral-like structures in the disc.
This is illustrated in Figure \ref{gasdisc} which shows the surface
density structure of the gaseous disc, once it has reached this
long-lived, self-gravitating state. Although it is not clear that
a protostellar disc could be self-gravitating at radii of a few au, 
it is quite likely that it plays a vital role in transporting angular
momentum in the earliest phases of star formation \citep{linpringle90}.
\citet{bellin94} also invoke self-gravity at $\sim 1$ au in their models
of FU Orionis outbursts, and the presence of a dead zone 
\citep{gammie96,armitage2001} could also allow a self-gravitating region to
extend to small radii. 

Once the gas disc has evolved into this self-gravitating state, we
introduce the `planetesimal' disc as discussed in \S \ref{numsims}.
For the gas disc we consider here, the largest radial drift rates will
occur for particles with sizes of $\sim 100$ cm (see \citealt{rice04}).
It is particles of this size that we expect to become the most strongly
concentrated in the spiral structures \citep{haghighipour03a, rice04}.
Particles significantly larger, or significantly smaller, would not be
as strongly influenced by the spiral structures.

\begin{figure}
\centerline{\psfig{figure=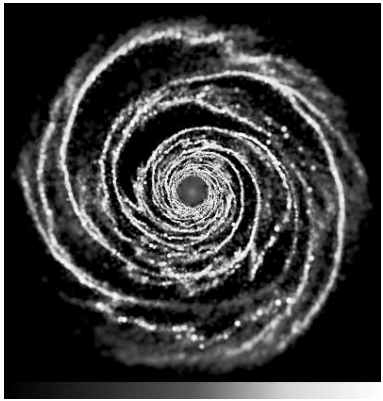, width=50mm}}
\caption{Surface density structure of the `planetesimal' disc
consisting of $150$ cm particles $\sim 80$ years after the
`planetesimal' disc is introduced into the simulation. The scale shows
the logarithm of the surface density, $\Sigma_{\rm pl}$, with the scale
covering $-0.6 < {\rm log}(\Sigma_{\rm pl} / {\rm g cm}^{-2}) <
2.4$. In this simulation, the spiral structures in the `planetesimal'
disc have fragmented producing bound clumps.}
\label{rd150}
\end{figure}

We consider initially $150$ cm particles with a surface density $100$
times smaller than that of the gas disc.  Figure \ref{rd150} shows the
surface density structure of this disc $\sim 80$ years after these
particles are introduced. The colour scale differs from that in Figure
\ref{gasdisc} due to the different initial surface densities.  What is
clear from Figure \ref{rd150} is that the spiral structures have
fragmented into clumps. These clumps are bound and have densities
significantly larger than the that of the disc gas. We also find fragmentation
throughout the planetesimal disc.  At the time shown in Figure \ref{rd150}, the 
innermost clump is at $r = 2.1$ and the outermost is at $r = 16.2$. It is
likely that the outermost regions of the `planetesimal' disc would also
produce clumps if we were to run the simulation further.

\begin{figure}
\centerline{\psfig{figure=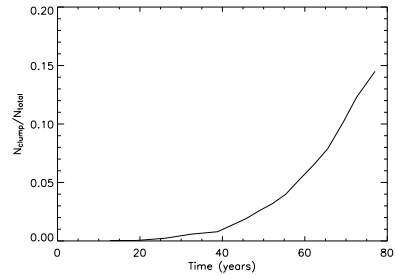, width=55mm}}
\caption{Fraction of `planetesimal' particles located in clumps with
densities exceeding $6 \times 10^{-8}$ g cm$^{-3}$. The first sign of
clump formation occurs about $20$ years after the solid particles are
introduced into the simulation with about $15 \%$ of the 'planetesimal'
particles located in dense clumps after $\sim 80$ years.}
\label{clumpfrac}
\end{figure}

The clump formation starts extremely quickly.  Figure \ref{clumpfrac}
shows the fraction of solid particles that have a local density
exceeding $6 \times 10^{-8}$ g cm$^{-3}$. This threshold density
exceeds the gas density everywhere in the disc and so ensures that
these particles are all located in dense clumps.  The first evidence of
clump formation occurs $\sim 20$ years after the solid particles are
introduced into the simulation, and almost $15 \%$ of the
`planetesimal' particles are located in dense clumps after $\sim 80$
years.  Although this gives some idea of the efficiency of the process,
we were unable to continue the simulation much further due to the
exceedingly small timesteps at which particles in dense clumps have to
be advanced. We therefore cannot assess if the clump fraction levels
off or continues to increase. The efficiency will also probably depend
on the structures within the gas disc.  A more massive disc, that is
likely to excite lower order modes, will probably have a different
efficiency to a lower mass disc that excites higher order modes
\citep{laughlin96, LR04}.

The above result shows that the particles that become strongly
concentrated in the self-gravitating, gaseous, spiral structures, may
achieve densities that could lead to planetesimal formation through
direct gravitational collapse. To test that this is not simply a
numerical effect, we carried out some additional simulations. We
repeated the above simulation changing the gravitational softening for
the solids to $10^{-1}$ au and $10^{-3}$ au.  
To a certain extent, softening acts to effectively smooth out the mass
distribution.  If the softening is too large, this can act to suppress
clumping.  If it is too small, the `graininess' of the mass distribution
can exacerbate the gravitational dynamics, increasing the velocity disperion
of the particles.  In our simulations, the particle separation at the
beginning of clump formation was $\sim 10^{-2}$ au, comparable to our original
softening of $10^{-2}$ au. Although clumps
formed for all of the softenings considered, the maximum clump density 
was much lower when the largest softening was used, consistent with clump
formation being suppressed for large softenings. For the smallest softening 
considered, clump formation was delayed relative to the larger softenings.  
This delay in fragmentation is, as mentioned earlier, a consequence of the
increased velocity dispersion of the solid particles due to
softening not being large enough to effectively reduce the `graininess' 
of the particle distribution. We also
performed a simulation with $1500$ cm particles, also with an initial
surface density $100$ times smaller than that of the disc gas, and
another with $150$ cm particles, but with an initial surface density
$1000$ times smaller than that of the disc gas.  In both of these
simulations the gravitational softening was $10^{-2}$ au.

\begin{figure}
\centerline{\psfig{figure=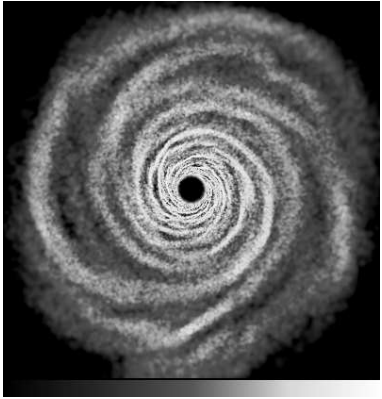, width=50mm}}
\caption{Surface density structure of the `planetesimal' disc
consisting of $1500$ cm particles $\sim 100$ years after the
`planetesimal' disc is introduced into the simulation. The scale shows
the logarithm of the surface density, $\Sigma_{\rm pl}$, with the scale
covering $-0.6 < {\rm log}(\Sigma / {\rm g cm}^{-2}) < 2.4$.}
\label{rd1500}
\end{figure}

Figure \ref{rd1500} shows the surface density structure of the
simulation with $1500$ cm particles.  As illustrated by \citet{rice04}
particles of this size are largely decoupled from the disc gas, are
essentially unaffected by gas drag, and hence do not become
significantly concentrated in the gaseous spiral structures.  The
spiral structures in the `planetesimal' disc effectively match that of
the gas disc and are produced primarily by the gravitational potential
of the gaseous disc, and not by gas drag.

\begin{figure}
\centerline{\psfig{figure=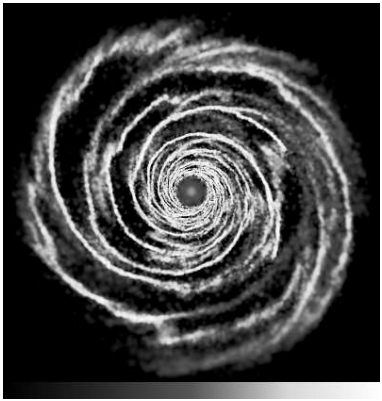, width=50mm}}
\caption{Surface density structure of the `planetesimal' disc
consisting of $150$ cm particles $\sim 80$ years after the
`planetesimal' disc is introduced into the simulation. The scale shows
the logarithm of the surface density, $\Sigma_{\rm pl}$, with the scale
covering $-1.6 < {\rm log}(\Sigma_{\rm pl} / {\rm g cm}^{-2}) < 1.4$.}
\label{rd150_m_10}
\end{figure}

Figure \ref{rd150_m_10} shows the surface density structure of the
simulation with $150$ cm particles, with an initial surface density
$1000$ times smaller than that of the gas disc.  In this case, the
particles do become concentrated in the gaseous spiral structures,
producing very narrow spiral structures in the `planetesimal' disc.
However, unlike the simulation with the higher initial surface density,
the densities achieved do not, in this simulation with the lower
initial surface density, lead to fragmentation and clump formation in
the `planetesimal' disc.  It appears, therefore, that the gravitational
collapse of solids requires particles that will become strongly
concentrated in the spiral density and that have a sufficiently high
initial surface density, relative to the surface density in the gas
disc.

\section{Discussion and conclusions}
The simulations presented here extends the work of \citet{rice04} which
showed that particles of a particular size (in their case $100$ cm
particles), could become very strongly concentrated in the spiral
structures present in a quasi-steady, self-gravitating gaseous
protoplanetary disc.

In this work we show that if the initial `planetesimal' surface density
is sufficiently high, the `planetesimal' disc may undergo fragmentation
to produce bound objects. It is this process that may provide a
mechanism for producing kilometre-sized and larger planetesimals.  In
the simulations here each bound fragment consists of $\sim 100$
pseudo-particles and hence have masses of $\sim 0.5$ M$_{\rm
earth}$. These clumps are extremely strongly bound with the magnitude
of the gravitational potential energy many times greater than the
kinetic energy.  It is therefore likely that each bound clump may
represent a number of smaller fragments, rather than a single bound
object.

We are therefore not suggesting that this process would lead, immediately,
to earth mass-like objects, but simply that the densities and velocity 
dispersions of the particles concentrated in the self-gravitating spiral
can lead to fragmentation of the `planetesimal' disc.

We should stress, however, that the process here differs considerably
from the standard \citet{goldreichward} mechanism (see also
\citealt{youdin05}) which requires a low velocity dispersion for
fragmentation of the `planetesimal' disc.  In our simulations,
fragmentation occurs when the velocity dispersion of the particles is
comparable to the sound speed of the disc gas.  It is the enhanced
density that leads to the fragmentation, rather than the reduced
velocity dispersion.

It does appear, however, that for fragmentation to occur, the surface
density of particles in the appropriate
 size range must be relatively
high.  This could occur if, while the disc is still massive, particle
growth up to $\sim$ metre sizes occurs through collisional growth.
Growth beyond metre sizes may, however, be difficult due to the
sticking efficiency becoming relatively small \citep{supulver97} and
there would therefore be a pile up of these particles.  Once the
surface density of these particles reaches a critical value, subsequent
growth would then occur through gravitational collapse triggered by the
density of these particles being enhanced in the spiral structures of
the gaseous disc. Additionally, the dependence of this process on the
solid to gas ratio is consistent with planet formation being more
likely in higher metallicity environments \citep{santos04, fischer05}.

The process described here potentially solves a major problem in the
standard planet formation scenario.  Rather than rapidly migrating into
the central star, $\sim$ metre-sized particles become concentrated in
self-gravitating spiral structures.  The densities achieved can then
lead to planetesimal formation via direct gravitational collapse.  In
this scenario, kilometre-sized planetesimals form very early, removing
a major bottleneck in the planet formation process.
\section*{acknowledgements}
The computations reported here were performed using the UK
Astrophysical Fluids Facility (UKAFF), and Datastar at the San Diego
Supercomputing Center (SDSC). This work was supported by NASA under
grants NAG5-13207 and NNG04GL01G from the Origins of Solar Systems and
Astrophysics Theory Programs, and by the NSF under grant
AST~0407040. The authors thank Matthew Bate for usefule discussions. 

\bibliographystyle{mn2e} 

\bibliography{lodato}

\end{document}